\documentclass{article}

\usepackage{graphicx}

 \usepackage[creativeai, final]{neurips_2025}

\usepackage[utf8]{inputenc} 
\usepackage[T1]{fontenc}    
\usepackage{hyperref}       
\usepackage{url}            
\usepackage{booktabs}       
\usepackage{amsfonts}       
\usepackage{nicefrac}       
\usepackage{microtype}      
\usepackage[table]{xcolor}

\usepackage{tabularx}
\usepackage{booktabs}
\usepackage{multirow}
\usepackage[table]{xcolor}
\usepackage{adjustbox}
\usepackage{amsmath}

\definecolor{LightBlue}{RGB}{100,150,255}
\hypersetup{
    colorlinks=true,
    linkcolor=black,
    citecolor=LightBlue,
    urlcolor=LightBlue,
}

\title{Creative Synthesis of Kinematic Mechanisms}

\author{%
  Jiong Lin $^{*}$
  \quad Jialong Ning 
  \quad Judah Goldfeder 
  \quad Hod Lipson
  \thanks{Corresponding authors: \href{jl6017@columbia.edu}{\color{black}{jl6017@columbia.edu}}, \href{hod.lipson@columbia.edu}{\color{black}{hod.lipson@columbia.edu}}. }
  \\
  Creative Machines Lab, Columbia University, 
  New York, NY 10027\\
  \href{https://jl6017.github.io/GenMech/}{https://jl6017.github.io/GenMech/}
}

\begin{document}

\maketitle
\begin{abstract}
In this paper, we formulate the problem of kinematic synthesis for planar linkages as a cross-domain image generation task. We develop a planar linkages dataset using RGB image representations, covering a range of mechanisms: from simple types such as crank-rocker and crank-slider to more complex eight-bar linkages like Jansen’s mechanism. A shared-latent variational autoencoder (VAE) is employed to explore the potential of image generative models for synthesizing unseen motion curves and simulating novel kinematics. By encoding the drawing speed of trajectory points as color gradients, the same architecture also supports kinematic synthesis conditioned on both trajectory shape and velocity profiles. 
We validate our method on three datasets of increasing complexity: a standard four-bar linkage set, a mixed set of four-bar and crank-slider mechanisms, and a complex set including multi-loop mechanisms.
Preliminary results demonstrate the effectiveness of image-based representations for generative mechanical design, showing that mechanisms with revolute and prismatic joints, and potentially cams and gears, can be represented and synthesized within a unified image generation framework.
Code and dataset are available at: \href{https://github.com/jl6017/GenMech}{GitHub}, \href{https://huggingface.co/datasets/jl6017/GenMech}{Hugging Face}.

\end{abstract}

\section{Introduction}
\begin{figure}[!t]
    \centering
    \includegraphics[width=1\linewidth]{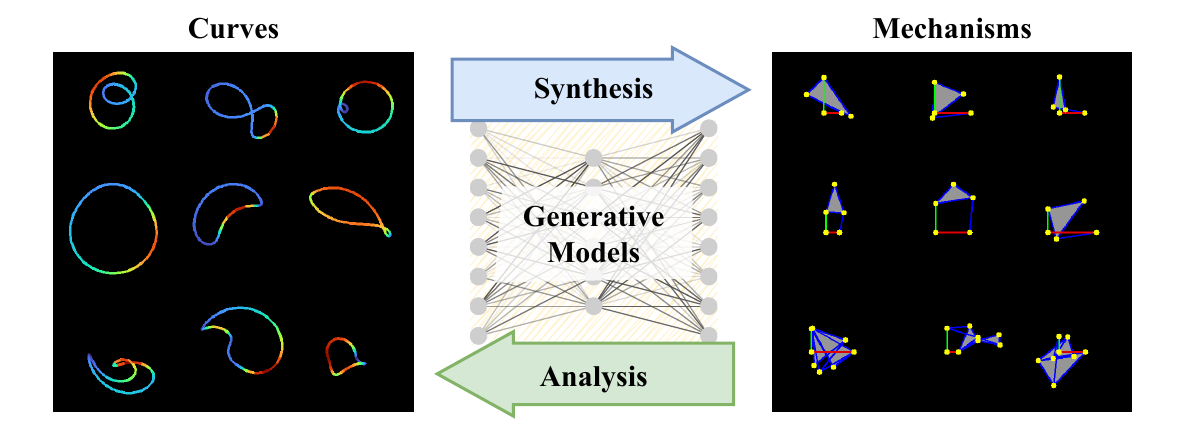}
    \caption{We propose an image based generative framework for kinematic synthesis and analysis. The model supports both synthesis and analysis using the same architecture and representation.  Trajectory speed is encoded using color gradients (left), while types of links and joints are encoded using predefined colors (right).
    }
    \label{fig:intro}
    \vspace{-1em}
\end{figure}

Kinematic synthesis is a long-standing problem in mechanical engineering. The objective is to design mechanisms that can generate a desired motion trajectory, using mechanical components such as links, sliders, gears, and cams. Within this broad area, the synthesis of planar linkages represents an important but still challenging subproblem. 
Classic examples such as the Watt's and Stephenson's linkages \cite{luo2018reconfigurable, bai2016unified}, as well as more complex designs like Jansen’s mechanism \cite{nansai2015jansen}, illustrate how carefully designed linkages can produce sophisticated and purposeful motion. However, the synthesis problem remains challenging due to the complex and irregular nature of the design space. This complexity arises from the need to ensure a specific degree of freedom, manage redundant components, avoid motion singularities \cite{maloisel2021singularity}, and account for multiple mechanisms that may produce the same trajectories \cite{zarkandi2021novel}. These challenges span both the \textit{analysis} perspective, simulating trajectories from given mechanisms, and \textit{synthesis} perspective, generating mechanisms from motion curves. We demonstrate these symmetric processes in Figure~\ref{fig:intro}.

Analytical approaches to kinematic synthesis date back to the 19th century, most notably with Burmester theory \cite{ceccarelli2006burmester}, which provides closed-form geometric solutions for linkage design under discrete pose constraints. With the advent of computational methods, optimization-based techniques such as evolutionary algorithms \cite{lipson2006relaxation, lipson2008evolutionary} and interactive design tools like LinkEdit \cite{bacher2015linkedit} have enabled the synthesis of more complex and non-intuitive mechanisms. More recently, the availability of large-scale datasets \cite{heyrani2022links, nurizada2025dataset} has enabled learning-based methods that train neural networks, including fully connected architectures, VAEs \cite{higgins2017beta, deshpande2019computational, nurizada2025path}, and multi-modal encoders like CLIP \cite{radford2021learning, nobari2024link}, to predict mechanisms from motion examples. However, existing methods often rely on task-specific data structures such as graphs or joint coordinate lists, which restrict generalization across different types of mechanisms. In contrast, image-based representations provide a unified and scalable format that naturally encodes both motion and structure, enabling the same model architecture and training pipeline to be applied across a wide variety of kinematic structures.

In this work, we propose a cross domain generative framework that jointly models mechanical structures and their motion trajectories. Trained on a curated dataset of paired images of mechanisms and curves, the model learns a shared latent representation that bridges the two domains. Once trained, it enables both kinematic synthesis and analysis.
Our contributions are as follows:
1). \textbf{A new dataset} of paired RGB images of planar mechanisms and their corresponding motion trajectories, covering diverse linkage types.
2). \textbf{An end-to-end image generative model} that learns a shared latent representation across the mechanism and trajectory domains.
3). \textbf{Demonstration of cross domain synthesis and analysis}, enabling high fidelity translation from trajectory to mechanism and from mechanism to trajectory.

\section{Related Work}
\begin{table}[!htbp]
\centering
{
\scriptsize
\rowcolors{2}{gray!10}{white}
\begin{tabularx}{\textwidth}{X c c c c}
\toprule
\textbf{Paper} & \textbf{Dataset} & \textbf{Curve Rep.} & \textbf{Mechanism Rep.} & \textbf{Method} \\
\midrule
Lipson \cite{lipson2008evolutionary} (2008) & Not fixed & Coordinates &  Operator tree & Genetic Programming \\
Vermeer et al.\cite{vermeer2018kinematic} (2018) & Not fixed & Coordinates  &  Operator tree & Reinforcement Learning \\
Deshpande et al.\cite{deshpande2021image} (2021) & 4-links and 6-links & Image & Coordinates & VAE+KNN \\
Pan et al.\cite{pan2023joint} (2023) & Not fixed & Coordinates  & Graph & Optimization \\
Fogelson et al.\cite{fogelson2023gcp} (2023) & Max 16-links & Coordinates & Graph & GCN+Reinforcement Learning  \\
Nobari et al.\cite{heyrani2022links, nobari2024link} (2024) & Max 20-links & Coordinates  & Graph & CLIP+BFGS \\
Lee et al.\cite{lee2024deep} (2024) & 4-links & Image  & Graph & cGAN \\
Nurizada et al.\cite{nurizada2025path} (2025) & 4-links \& sliders & Image & Coordinates & $\beta$-VAE \\
Ours & Max 12-links \& sliders & Image \& Velocities  & Image or Video & Shared-latent VAEs \\
\bottomrule
\end{tabularx}
}
\caption{Summary of selected relevant kinematic synthesis papers.}
\label{tab:kinematic_synthesis}
\vspace{-1em} 
\end{table}
{\bf{Kinematic synthesis.}} 
Early work attempted to solve kinematic synthesis by interpolation. Simulation was used to generate a large database of mechanisms, and a neural network guided synthesis by interpolating a new mechanism from similar ones in the database~\cite{vasiliu2001dimensional}. Other approaches used a tree-based mechanism representation, relying on genetic programming~\cite{lipson2006relaxation} or reinforcement learning~\cite{fogelson2023gcp} to search the space of trees and find mechanisms with the desired properties. 

Recent work has mostly focused on deep learning. A variety of generative models have been explored, including cGANs~\cite{lee2024deep} and VAEs~\cite{nurizada2025path}. Contrastive Learning has also been applied to enable rapid retrieval from massive mechanism databases~\cite{ nobari2024link}. The rise of data driven approaches has further spurred the release of several high quality datasets
~\cite{nurizada2025dataset,heyrani2022links} and even attempts at 3D synthesis~\cite{cheng2022exact}.
Related to synthesis, mechanism design tools have also been explored. One example is mechanism editing, whereby a user can interact with an existing mechanism, and the solver allows for the user to add new properties while specifying properties that need to be preserved~\cite{bacher2015linkedit}. Another tool automated conversion from hand drawn sketches to digital representations~\cite{nurizada2024transforming}. Yet another presented an interactive design system for the creation of mechanical characters~\cite{coros2013computational} and linkage based characters~\cite{thomaszewski2014computational}, and multi-stable structures~\cite{zhang2021computational}.

{\bf{Generative models }}have progressed from Variational Autoencoders (VAEs)~\cite{kingma2013auto}, which learn probabilistic latent representations, to Diffusion Models such as DDPMs~\cite{ho2020denoising}, which iteratively transform noise into data through denoising steps. Latent Diffusion Models (LDMs)~\cite{rombach2022high} perform diffusion in a compressed latent space, while Flow Matching~\cite{lipman2022flow} offers a continuous-time formulation that learns transformation between distributions.
Cross-domain generative models aim to translate data from one domain to another while preserving semantic consistency. Applications includes image-to-image translation~\cite{taigman2017unsupervised, kim2017learning}, few-shot domain adaptation~\cite{ojha2021few}, and cross-modal 3D synthesis~\cite{liu2023zero, long2024wonder3d}. 
Multi-modal generative models extend this idea to learning joint representations across different modalities, such as video and audio~\cite{ngiam2011multimodal}.
By jointly modeling multiple domains or modalities, these methods enable richer synthesis capabilities and bidirectional translation between diverse input types.
Previous works have primarily focused on datasets such as SVHN to MNIST~\cite{hoffman2018cycada} or style translation~\cite{richardson2021encoding} between domains. In this work, we introduce kinematic synthesis as a new cross-domain generative modeling task, well-defined and application-relevant, and present a data generation pipeline with preliminary results on three datasets using the shared-latent VAEs.

\section{Approach}

\begin{figure}[!t]
    \centering
    \includegraphics[width=1\linewidth]{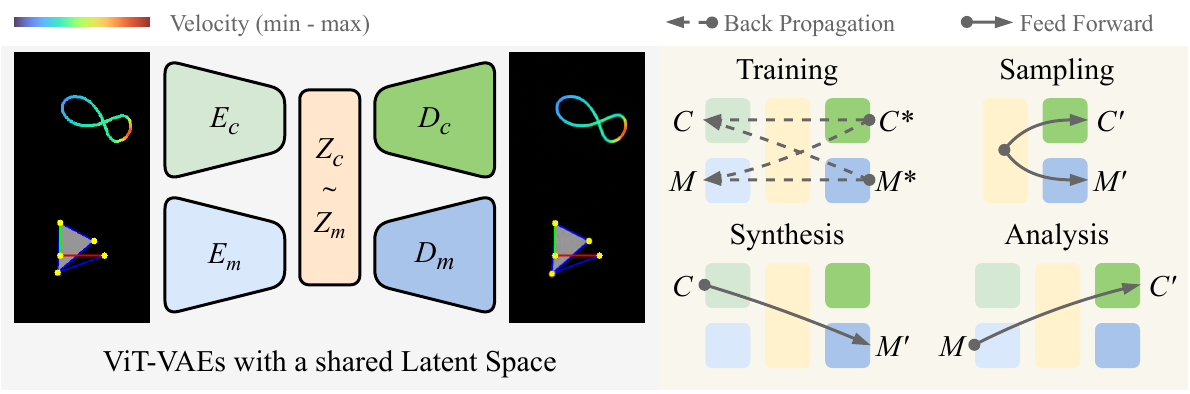}
    \caption{
    Overview of our shared-latent VAE framework for cross-domain kinematic synthesis and analysis. Curve and mechanism images ($C$, $M$) are encoded into latent embeddings ($Z_c$, $Z_m$), which are aligned in a shared latent space. During training, both reconstruction ($\hat{C}$, $\hat{M}$) and cross-domain prediction ($\hat{C}_m$, $\hat{M}_c$) are supervised via back-propagation. At inference time, the model enables synthesis (from $C$ to $M'$) and analysis (from $M$ to $C'$) through feed-forward decoding. Following MAE~\cite{he2022masked}, we adopt an asymmetric ViT encoder–decoder design.
    }
    \label{fig:approach}
    \vspace{-1em}
\end{figure}
\vspace{-0.5em}  
\subsection{Shared-latent VAEs}
Overview of our method shown in Figure~\ref{fig:approach}.
Let $C$ and $M$ denote the input curve and mechanism images. $E_c$, $E_m$ are the encoders, and $D_c$, $D_m$ are the decoders for the curve and mechanism branches. The latent codes are $Z_c = E_c(C)$ and $Z_m = E_m(M)$, sampled from a shared latent distribution $Z \sim \mathcal{N}(0, I)$. The reconstructions are $\hat{C} = D_c(Z_c)$ and $\hat{M} = D_m(Z_m)$, while the cross-domain predictions are $\hat{M}_c = D_m(Z_c)$ and $\hat{C}_m = D_c(Z_m)$.
The training loss is defined as:
\begin{align}
\mathcal{L}_{\text{Shared-latent-VAEs}} =\ & \| C - \hat{C} \|^2 + \| M - \hat{M} \|^2 \nonumber 
 + \beta \cdot \left[ \text{KL}(q(Z_c|C) \| p(Z)) + \text{KL}(q(Z_m|M) \| p(Z)) \right] \nonumber \\
& + \lambda \cdot \| Z_c - Z_m \|^2 \nonumber + \gamma \cdot ( \| \hat{M}_c - M \|^2 + \| \hat{C}_m - C \|^2 )
\end{align}

The loss function consists of five components. The \textbf{reconstruction loss} measures the image reconstruction error for both the curve and mechanism domains, ensuring that each encoder-decoder pair can accurately reproduce its input. The \textbf{KL divergence} term regularizes the latent codes $Z_c$ and $Z_m$ to follow a standard normal distribution, promoting smoothness and structure in the latent space. The \textbf{latent similarity loss} encourages the curve and mechanism embeddings to be close, facilitating a shared latent representation across domains. Lastly, the \textbf{cross-domain prediction loss} ensures that the latent codes contain sufficient information to decode meaningful counterparts in the other domain, supporting both synthesis and analysis tasks.

\subsection{Dataset}
\begin{figure}[!t]
    \centering
    \includegraphics[width=1\linewidth]{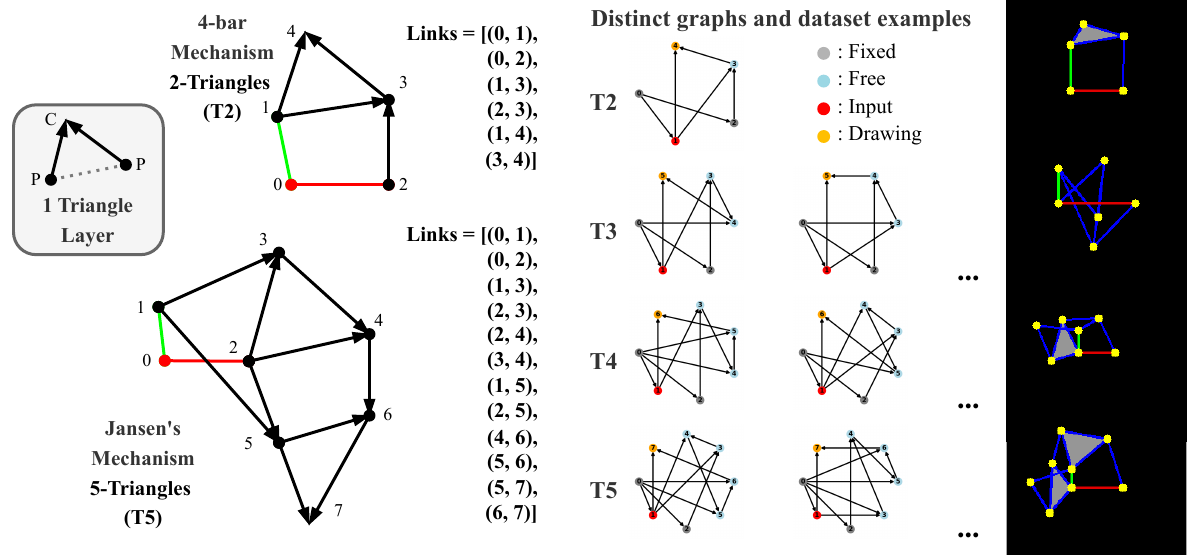}
    \caption{Examples of dataset construction. \textbf{Left:} Example mechanisms (4-bar mechanism and Jansen’s linkage) built with 2 and 5 triangle layers, with their corresponding sequences of link connections. \textbf{Right:} For each complexity level (T2–T5), two representative graphs from different isomorphism classes are shown, along with rendered mechanism examples from the dataset.}
    \label{fig:dataset}
    \vspace{-1em}
\end{figure}
We construct a dataset of 1-DOF planar mechanisms by recursively applying triangle-based operators, starting from two fixed joints and one input joint, as shown in Figure~\ref{fig:dataset}. 
There are multiple operators to preserve the 1-DOF property while expanding the mechanism, such as T-operator and D-operator~\cite{lipson2008evolutionary}.
Following recent work, we use the J-operator from the LINKS dataset~\cite{heyrani2022links} and the recurrent triangle solver introduced in LinkEdit~\cite{bacher2015linkedit} to build mechanisms stacking triangle layers. Our goal is to include historically significant mechanisms such as Watt's, Stephenson's, and Jansen's, all of which share key properties: they begin with two fixed nodes, have one drawing node, and can be constructed with triangle layers.~\footnote{Sub-variants such as Watt-II and Stephenson-III, which involve three fixed nodes, are not included in the current version dataset. The dataset can be extended to include these cases by adjusting the filtering parameters.}

To generate a compact yet expressive dataset spanning from simple four-bar linkages to complex mechanisms such as Jansen’s, we start with all possible combinations and then apply a set of structural filters. We begin with three initial nodes and recursively add one node at each layer by selecting two existing nodes as parents. This results in a total of $\prod_{i=3}^{k+2} \binom{i}{2}$ combinations for $k$ triangle layers, which corresponds to the first row in Table~\ref{tab:triangle_filters_full}.
For example, when \( k = 2 \), the number of combinations is 
\( \binom{3}{2} \cdot \binom{4}{2} = 18 \); 
when \( k = 5 \), \( \binom{3}{2} \cdot \binom{4}{2} \cdot \binom{5}{2} \cdot \binom{6}{2} \cdot \binom{7}{2} 
= 56{,}700 \).
We only retain mechanisms that satisfy the following conditions: 1) a single drawing node, since mechanisms with multiple output points can be described by simpler substructures; 2) no redundant triangles, as adding a triangle onto an already rigid structure does not increase structural variety; 3) exactly two fixed nodes, to match the properties of the target mechanisms and avoid introducing additional grounded points; and 4) one representative graph per isomorphic class, removing any structure that differs only by the sequence of construction. As a result, we include 396 distinct linkage structures, along with an equal number of crank-slider-based structures obtained by changing the starting point of graph construction.

\begin{table}[!htbp]
\centering
{\scriptsize
\rowcolors{2}{gray!10}{white}
\begin{tabularx}{0.65\textwidth}{X c c c c c c c}
\toprule
\textbf{Filters | Triangle layers} & \textbf{T0} & \textbf{T1} & \textbf{T2} & \textbf{T3} & \textbf{T4} & \textbf{T5} & \textbf{Total} \\
\midrule
Initial: All combinations   & 1     & 3     & 18    & 180   & 2700  & 56700 & 59602 \\
Filter 1: One drawing node  & 0     & 1     & 5     & 31    & 257   & 2803  & 3097 \\
Filter 2: Two fixed nodes  & 0     & 0     & 1     & 11    & 107   & 1227  & 1346 \\
Filter 3: No redundant links  & 0     & 0     & 1     & 8     & 68    & 632   & 709 \\
Filter 4: Isomorphic graphs  & 0     & 0     & 1     & 7     & 47    & 341   & 396 \\
\bottomrule
\end{tabularx}
\vspace{0.5em}  
}
\caption{Number of graphs retained after applying each structural filter, for the number of triangle layers from 0 (T0) to 5 (T5).}
\label{tab:triangle_filters_full}
\vspace{-1em} 
\end{table}

Color encoding is applied to both curve and mechanism images. For curves, the drawing-end speed is mapped to a color scale. For mechanisms, each component is assigned a predefined color: the base link is red, the input link is green, other links are blue, and joints are yellow. This encoding enables explicit validation when deriving parametric models.
Inspired by the design of the \textit{Strandbeest's}~\cite{nansai2015jansen} walking leg, we observe that flattened contact points and steady horizontal motion of the output nodes promote stable forward movement and long locomotion overlap. Incorporating speed information into kinematic synthesis is both beneficial and easily integrated in RGB image representations.

\section{Experiments}
\begin{figure}[!t]
    \centering
    \includegraphics[width=1\linewidth]{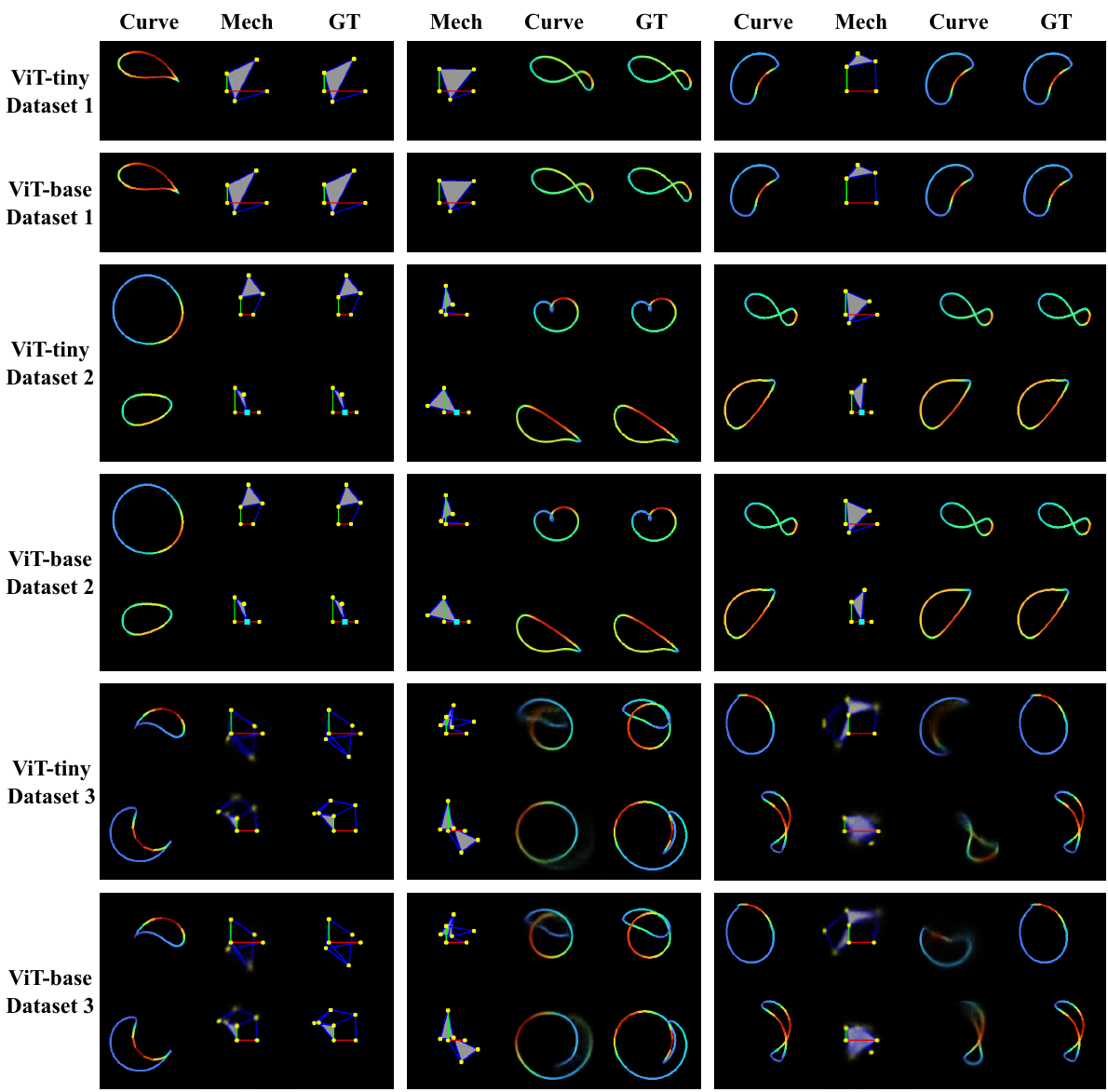}
    \caption{Qualitative results. Columns: (1) Curve to synthesized mechanism, compared with ground-truth mechanism; (2) Mechanism to calculated curve, shown beside the ground-truth curve; (3) Curve-to-mechanism-to-curve, compared with ground-truth curve. Rows list ViT-tiny and ViT-base~\cite{dosovitskiy2020image} models on three datasets.}
    \label{fig:result}
    \vspace{-1em}
\end{figure}
We open-source the data generation code and provide three example datasets used to validate the proposed shared-latent VAE for image-based synthesis and analysis. \textbf{Dataset-1} contains 100K four-bar mechanisms, including both crank–rocker and double–crank cases, referred to as T2. \textbf{Dataset-2} contains 200K mixed four-bar mechanisms and crank–slider mechanisms, referred to as T2ST2. \textbf{Dataset-3}, denoted as T4, is a subset of our complete collection of distinct graphs, comprising 55 graphs up to T4, with 10K samples collected for each graph. We repeated the data collection process and used the resulting test set for all three experiments.

\begin{table}[!t]
\centering
{\scriptsize
\begin{tabular}{
  p{0.17\textwidth}|
  >{\centering\arraybackslash}p{0.1\textwidth}
  >{\centering\arraybackslash}p{0.1\textwidth}
  >{\centering\arraybackslash}p{0.1\textwidth}|
  >{\centering\arraybackslash}p{0.1\textwidth}
  >{\centering\arraybackslash}p{0.1\textwidth}
  >{\centering\arraybackslash}p{0.1\textwidth}
}
\toprule
\multirow{3}{*}{\textbf{Models}} 
& \multicolumn{3}{c|}{\textbf{Dataset 1 (T2) {\textbf{PSNR} $\uparrow$}}} 
& \multicolumn{3}{c}{\textbf{Dataset 2 (T2 \& ST2) {\textbf{PSNR} $\uparrow$}}} \\
\cmidrule(lr){2-4} \cmidrule(lr){5-7}
& \textbf{Synthesis} & \textbf{Analysis} & \textbf{C2M2C}
& \textbf{Synthesis} & \textbf{Analysis} & \textbf{C2M2C} \\
\midrule
ViT-tiny     & 27.63 ± 2.28 & 26.59 ± 2.21 & 26.17 ± 2.36 & 29.48 ± 3.27 & 27.97 ± 2.60 & 28.01 ± 2.72 \\
ResNet-18     & \textbf{28.58} ± 3.15 & 26.83 ± 2.61 & 27.03 ± 2.43 & 30.83 ± 4.54 & 28.20 ± 3.00 & 27.82 ± 3.34 \\
ViT-base     & 28.54 ± 3.26 & \textbf{26.90} ± 2.43 & \textbf{27.18} ± 2.22 & \textbf{30.87} ± 4.76 & \textbf{28.43} ± 3.07 & \textbf{28.40} ± 3.20 \\
\bottomrule
\end{tabular}
\vspace{0.5em}  
}
\caption{\textbf{Baseline comparison.} PSNR (mean ± std) for image accuracy in synthesis, analysis, and curve-to-mechanism-to-curve (C2M2C) tasks, on two datasets for three models~\cite{dosovitskiy2020image, he2016deep}.}
\label{tab:psnr_baseline}
\vspace{-1em}
\end{table}

\begin{table}[!t]
\centering
{\scriptsize
\begin{tabular}{
  p{0.17\textwidth}|
  >{\centering\arraybackslash}p{0.1\textwidth}
  >{\centering\arraybackslash}p{0.1\textwidth}
  >{\centering\arraybackslash}p{0.1\textwidth}|
  >{\centering\arraybackslash}p{0.1\textwidth}
  >{\centering\arraybackslash}p{0.1\textwidth}
  >{\centering\arraybackslash}p{0.1\textwidth}
}
\toprule
\multirow{3}{*}{\textbf{Ablations}} 
& \multicolumn{3}{c|}{\textbf{Dataset 1 (T2) \textbf{PSNR} $\uparrow$}} 
& \multicolumn{3}{c}{\textbf{Dataset 2 (T2 \& ST2) \textbf{PSNR} $\uparrow$}} \\
\cmidrule(lr){2-4} \cmidrule(lr){5-7}
& \textbf{Synthesis} & \textbf{Analysis} & \textbf{C2M2C}
& \textbf{Synthesis} & \textbf{Analysis} & \textbf{C2M2C} \\
\midrule
Black-white input      & 25.61 ± 3.10 & 23.42 ± 2.40 & 22.86 ± 2.35 & 27.38 ± 3.91 & 24.70 ± 2.83 & 24.22 ± 3.24 \\
w/o analysis loss     & 26.43 ± 1.87 & 12.50 ± 0.38 & 12.49 ± 0.38 & 29.03 ± 3.20 & 12.52 ± 0.33 & 12.52 ± 0.33 \\
w/o synthesis loss    & 12.41 ± 0.29 & 26.30 ± 1.89 & 18.55 ± 1.46 & 12.55 ± 0.30 & 27.01 ± 2.41 & 18.71 ± 1.28 \\
w/o cross-domain loss  & 18.21 ± 0.96 & 19.22 ± 1.17 & 18.63 ± 1.73 & 18.61 ± 1.26 & 18.91 ± 1.57 & 19.03 ± 1.56 \\
Complete loss (ViT-tiny)        & \textbf{27.63} ± 2.28 & \textbf{26.59} ± 2.21 & \textbf{26.17} ± 2.36 & \textbf{29.48} ± 3.27 & \textbf{27.97} ± 2.60 & \textbf{28.01} ± 2.72 \\
\bottomrule
\end{tabular}
\vspace{0.5em}  
}
\caption{\textbf{Ablation study.} Effect of input color and loss function terms on PSNR (mean ± std) for synthesis, analysis, and C2M2C tasks.}
\label{tab:psnr_ablation}
\vspace{-1em}
\end{table}

\begin{figure}[!t]
    \centering
    \includegraphics[width=1\linewidth]{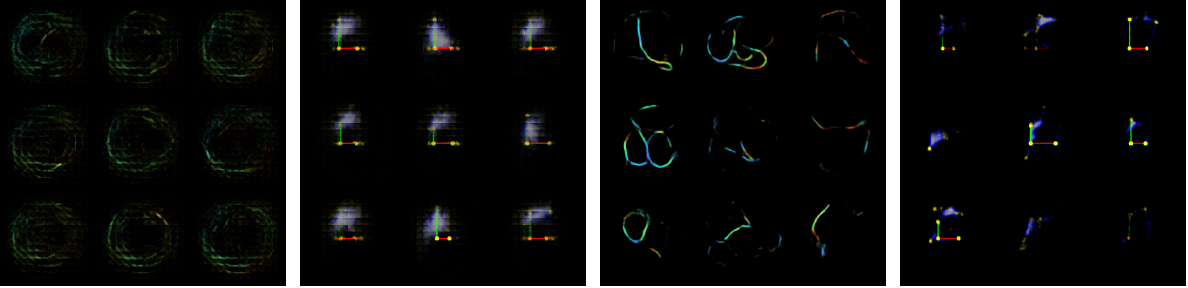}
    \caption{Samples from the latent space, shown in a $3\times 3$ grid: from left to right: ViT-base with $\beta=10$ (curves, mechanisms) and CNN with $\beta=1$ (curves, mechanisms).
}
    \label{fig:sample}
\vspace{-1em}
\end{figure}

\textbf{Qualitative results.} We validate our method on three tasks: curve-to-mechanism synthesis, mechanism-to-curve analysis, and a two-stage generation where synthesis is followed by analysis. Figure~\ref{fig:result} presents results for all three tasks across the three datasets. ViT-base generally produces more accurate outputs than ViT-tiny, particularly as the complexity of the mechanisms increases.

\textbf{Quantitative results.} Tables~\ref{tab:psnr_baseline} and \ref{tab:psnr_ablation} report PSNR for all tasks. ViT-base achieves the highest scores across datasets compared with ResNet and ViT-tiny, especially in two-stage generation. Ablations show that removing synthesis or cross-domain losses significantly reduces accuracy, and color input outperforms black-and-white. The complete loss configuration yields the best overall performance. 

\section{Discussion}
\textbf{Limitations and future work.}
Figure~\ref{fig:sample} shows latent space samples for curves and mechanisms using ViT-base with $\beta=10$ and ResNet with $\beta=1$. The current sampling quality is limited, with blurred outputs and low-quality edges. This suggests that the VAE structure with high-capacity decoders may not have learned a well-structured normal distribution in the latent space. 

In future work, large-scale datasets and more advanced generative models are encouraged for this task. In our data generation pipeline, we provide the optional recordings of joint coordinates and videos of one period's mechanism motion, enabling future exploration of image-to-graph and image-to-video generation. Image-based representations for complex mechanism structures suffer from occlusion and ambiguity, whereas video representations can provide richer temporal and structural information.

\textbf{Conclusion.}
In this paper, we introduced kinematic synthesis as a cross-domain generative modeling task and proposed a shared-latent VAE framework for translating between mechanisms and motion curves. We constructed a scalable dataset covering simple to complex planar mechanisms and demonstrated synthesis, analysis, and two-stage generation on multiple datasets. Preliminary results show that image-based representations enable a unified representation and end-to-end generative synthesis of kinematic mechanisms. This work lays the groundwork for applying generative modeling into mechanism design, with potential applications in robotics.

\textbf{Acknowledgments:}
This work was supported in part by the US National Science
Foundation AI Institute for Dynamical Systems (DynamicsAI.org)
(grant no. 2112085). The author thanks Jinchen Ruan for designing the hardware machine for the real-world validation.

{
\small
\bibliographystyle{abbrv}
\bibliography{main}
}

\appendix
\section{Real-World Validation}
\begin{figure}[!htbp]
    \centering
    \includegraphics[width=1\linewidth]{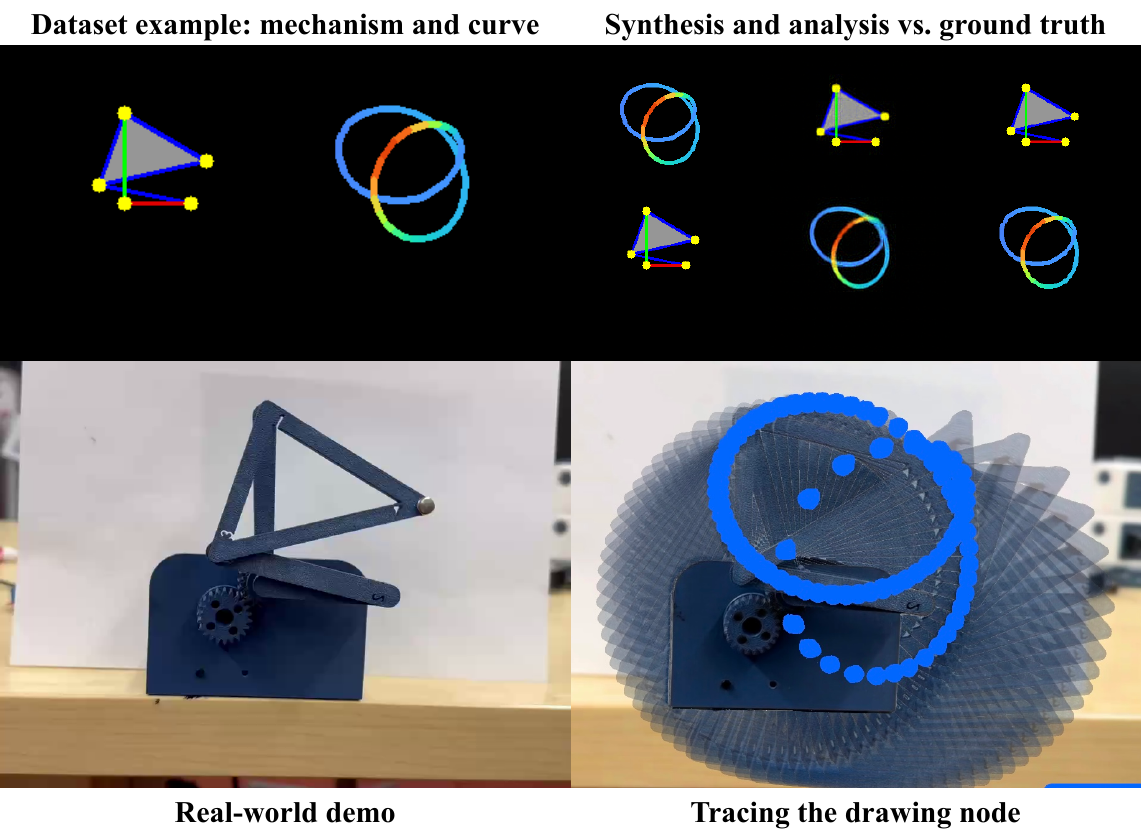}
    \caption{Examples of mechanism–curve pairs and real-world validation. Top left: a dataset example showing a mechanism and its corresponding drawing curve. Top right: synthesis (curve-to-mechanism) and analysis (mechanism-to-curve) results compared with ground truth, using the ViT-Tiny model. Bottom: a 3D-printed mechanism driven by a servo motor and the traced drawing trajectory of its end-effector.}

    \label{fig:real}
    \vspace{-1em}
\end{figure}

\newpage
\section*{NeurIPS Paper Checklist}

\begin{enumerate}

\item {\bf Claims}
    \item[] Question: Do the main claims made in the abstract and introduction accurately reflect the paper's contributions and scope?
    \item[] Answer: \answerYes{}
    \item[] Justification: The abstract and introduction clearly state the main contributions: dataset creation, shared-latent VAE framework, and bidirectional synthesis and simulation. Claims match the presented results.

\item {\bf Limitations}
    \item[] Question: Does the paper discuss the limitations of the work performed by the authors?
    \item[] Answer: \answerYes{}.
    \item[] Justification: Discussed in Section 5 Discussion: limitations in sampling quality, edge sharpness, and possible occlusion issues in image-based representation; suggestions for larger datasets and better generative models.

\item {\bf Theory assumptions and proofs}
    \item[] Question: For each theoretical result, does the paper provide the full set of assumptions and a complete (and correct) proof?
    \item[] Answer: \answerNA{}.
    \item[] Justification: The paper does not contain formal theorems or proofs

    \item {\bf Experimental result reproducibility}
    \item[] Question: Does the paper fully disclose all the information needed to reproduce the main experimental results of the paper to the extent that it affects the main claims and/or conclusions of the paper (regardless of whether the code and data are provided or not)?
    \item[] Answer: \answerYes{}.
    \item[] Justification: Datasets, code, and implementation details are provided (Abstract line 16, Section 4 Experiments). Hyperparameters and dataset construction steps are described.

\item {\bf Open access to data and code}
    \item[] Question: Does the paper provide open access to the data and code, with sufficient instructions to faithfully reproduce the main experimental results, as described in supplemental material?
    \item[] Answer: \answerYes{}.
    \item[] Justification: Links to GitHub and Hugging Face are included in the abstract; dataset generation code is open-sourced.

\item {\bf Experimental setting/details}
    \item[] Question: Does the paper specify all the training and test details (e.g., data splits, hyperparameters, how they were chosen, type of optimizer, etc.) necessary to understand the results?
    \item[] Answer: \answerYes{}.
    \item[] Justification: Training/test splits, dataset sizes, model architectures (ViT-tiny, ViT-base, ResNet-18), loss terms, and evaluation metrics (PSNR with mean±std) are specified in Sections 3–4.

\item {\bf Experiment statistical significance}
    \item[] Question: Does the paper report error bars suitably and correctly defined or other appropriate information about the statistical significance of the experiments?
    \item[] Answer: \answerYes{}.
    \item[] Justification: Results include mean and standard deviation for all quantitative metrics (Tables~\ref{tab:psnr_baseline} and~\ref{tab:psnr_ablation}).

\item {\bf Experiments compute resources}
    \item[] Question: For each experiment, does the paper provide sufficient information on the computer resources (type of compute workers, memory, time of execution) needed to reproduce the experiments?
    \item[] Answer: \answerYes{}.
    \item[] Justification: All experiments were conducted on a single NVIDIA RTX~5090 GPU. Training ViT-tiny on a 100k-sample dataset takes approximately 8 hours; ViT-base on the same dataset takes about 20 hours.
    
\item {\bf Code of ethics}
    \item[] Question: Does the research conducted in the paper conform, in every respect, with the NeurIPS Code of Ethics \url{https://neurips.cc/public/EthicsGuidelines}?
    \item[] Answer: \answerYes{}.
    \item[] Justification: The work uses synthetic data and does not involve human subjects, privacy-sensitive information, or unethical applications; it complies with the NeurIPS Code of Ethics.

\item {\bf Broader impacts}
    \item[] Question: Does the paper discuss both potential positive societal impacts and negative societal impacts of the work performed?
    \item[] Answer: \answerYes{}
    \item[] Justification: Positive impacts include advancing generative design for robotics and mechanical systems. Potential negative impacts involve unsafe or unreliable mechanisms if deployed without validation; open datasets are intended for research use only.
    
\item {\bf Safeguards}
    \item[] Question: Does the paper describe safeguards that have been put in place for responsible release of data or models that have a high risk for misuse (e.g., pretrained language models, image generators, or scraped datasets)?
    \item[] Answer: \answerNA{}
    \item[] Justification: The released models and datasets have low misuse potential and do not contain harmful or sensitive content.


\item {\bf Licenses for existing assets}
    \item[] Question: Are the creators or original owners of assets (e.g., code, data, models), used in the paper, properly credited and are the license and terms of use explicitly mentioned and properly respected?
    \item[] Answer: \answerYes{}
    \item[] Justification: External algorithms/tools (e.g., LINKS dataset, LinkEdit) are properly cited. 


\item {\bf New assets}
    \item[] Question: Are new assets introduced in the paper well documented and is the documentation provided alongside the assets?
    \item[] Answer: \answerYes{}
    \item[] Justification: A new dataset of paired RGB mechanism–curve images is documented in Section~3.2 and released along with the dataset generation code under the MIT License. Instructions for use and licensing are included in the repository.


\item {\bf Crowdsourcing and research with human subjects}
    \item[] Question: For crowdsourcing experiments and research with human subjects, does the paper include the full text of instructions given to participants and screenshots, if applicable, as well as details about compensation (if any)? 
    \item[] Answer: \answerNA{}
    \item[] Justification: The work does not involve human subjects or crowdsourced data collection.


\item {\bf Institutional review board (IRB) approvals or equivalent for research with human subjects}
    \item[] Question: Does the paper describe potential risks incurred by study participants, whether such risks were disclosed to the subjects, and whether Institutional Review Board (IRB) approvals (or an equivalent approval/review based on the requirements of your country or institution) were obtained?
    \item[] Answer: \answerNA{}
    \item[] Justification: Not applicable; no human subjects are involved.

\item {\bf Declaration of LLM usage}
    \item[] Question: Does the paper describe the usage of LLMs if it is an important, original, or non-standard component of the core methods in this research? Note that if the LLM is used only for writing, editing, or formatting purposes and does not impact the core methodology, scientific rigorousness, or originality of the research, declaration is not required.
    \item[] Answer: \answerNA{}
    \item[] Justification: The core method development does not involve LLMs as an original or non-standard component; LLMs were not used for the experiments or methodology.

\end{enumerate}

\end{document}